\documentclass[preprint]{aastex}
\usepackage{emulateapj5}
\usepackage{apjfonts}

\input psfig.tex

\def\aj{{AJ}}

\def\apj{{ApJ}}
\def\apjs{{ApJS}}

\def\flamb{erg s$^{-1}$ cm$^{-2}$ \AA$^{-1}$}

\def\ha{H$\alpha$}
\def\hb{H$\beta$}

\def\kms{km s$^{-1}$}
\def\lamb{$\lambda$}
\def\ll{$\lambda\lambda$}
\def\lax{{$\mathrel{\hbox{\rlap{\hbox{\lower4pt\hbox{$\sim$}}}\hbox{$<$}}}$}}
\def\gax{{$\mathrel{\hbox{\rlap{\hbox{\lower4pt\hbox{$\sim$}}}\hbox{$>$}}}$}}
\def\simlt{\lower.5ex\hbox{$\; \buildrel < \over \sim \;$}}
\def\simgt{\lower.5ex\hbox{$\; \buildrel > \over \sim \;$}}
\def\lum{erg s$^{-1}$}
\def\mbh{{$M_{\rm BH}$}}

\def\mnras{{MNRAS}}

\def\percm2{cm$^{-2}$}

\def\solmass{$M_\odot$}

\def\oiii{[\ion{O}{3}]}

\def\nii{[\ion{N}{2}]}

\def\whz{W Hz$^{-1}$}

\newdimen\digitwidth      
\setbox1=\hbox{0}       
\digitwidth=\wd1        
\catcode`"=\active      
\def"{\kern\digitwidth}

\slugcomment{ 
To appear in {\it The Astrophysical Journal Letters}}
\shorttitle{Young Radio Quasars}
\shortauthors{KIM ET AL.}

\begin{document}

\title{Evidence for AGN-driven Outflows in Young Radio Quasars}

\author{Minjin Kim\altaffilmark{1,2,3}, 
Luis C. Ho\altaffilmark{1},
Carol J. Lonsdale\altaffilmark{4},
Mark Lacy\altaffilmark{4},
Andrew W. Blain\altaffilmark{5},
and Amy E. Kimball\altaffilmark{4}}

\altaffiltext{1}{The Observatories of the Carnegie Institution for Science, 
813 Santa Barbara Street, Pasadena, CA 91101, USA}

\altaffiltext{2}{Korea Astronomy and Space Science Institute, Daejeon 305-348, 
Republic of Korea}

\altaffiltext{3}{KASI-Carnegie Fellow}

\altaffiltext{4}{National Radio Astronomy Observatory, 520 Edgemont Road, 
Charlottesville, VA 22903, USA}

\altaffiltext{5}{Physics and Astronomy, University of Leicester, Leicester, 
UK}

\begin{abstract}
We present near-infrared spectra of young radio quasars ($P_{1.4\, {\rm GHz}}
\approx 26 - 27$ \whz) selected from the
{\it Wide-Field Infrared Survey Explorer}. The detected objects have typical 
redshifts of $z \approx 1.6-2.5$ and bolometric luminosities $\sim 10^{47}$ 
\lum.  Based on the intensity ratios of narrow emission lines, we find that 
these objects are mainly powered by active galactic nuclei (AGNs), although 
star formation contribution cannot be completely ruled out. The host galaxies 
experience moderate levels of extinction, $A_V \approx 0-1.3$ mag.  The 
observed \oiii\ \lamb5007 luminosities and rest-frame $J$-band magnitudes 
constrain the black hole masses to lie in the range $\sim 10^{8.9}-10^{9.7}$ 
\solmass.  From the empirical correlation between black hole mass and host 
galaxy mass, we infer stellar masses of $\sim 10^{11.3}-10^{12.2}$ \solmass.  
The \oiii\ line is exceptionally broad, with full width at half maximum 
$\sim$1300 to 2100 \kms, significantly larger than that of ordinary distant 
quasars.  We argue that these large line widths can be explained by 
jet-induced outflows, as predicted by theoretical models of AGN feedback.
\end{abstract}


\section{Introduction}
Supermassive black holes (BHs) and their host galaxies appear to be closely 
linked in their formation and evolution (e.g., Magorrian et al. 1998; Gebhardt 
et al. 2000; Ferrarese \& Merritt 2000).  While the exact origin of their 
relationship is still under debate (Kormendy \& Ho 2013), theoretical models 
often invoke feedback from active galactic nuclei (AGNs) as a crucial 
mechanism for establishing the correlation.  
Outflows are believed to be associated with AGN feedback,
which regulate the growth of BHs and quench 
star formation by expelling gas (e.g., Di Matteo et al. 2005).  AGN feedback 
also appears to be required for cosmological models to explain the upper 
end of the luminosity function of galaxies (e.g., Bower et al. 2006).  Despite 
the potential importance of AGN feedback, direct observational evidence of 
this process is limited.  For example, extended \oiii\ emission in high-$z$ 
radio galaxies and ultraluminous infrared galaxies (ULIRGs) associated with 
AGNs are thought to be signs of AGN-driven outflows (e.g., Nesvadba 2007; 
Nesvadba et al. 2011; Harrison et al. 2012).  

Numerical simulations indicate that feedback is triggered during an intense and 
rapidly accelerated BH fueling phase during a gas-rich galaxy merger (Hopkins 
et al. 2008; Narayanan et al. 2010). Objects in the early feedback phase are 
likely to be both highly luminous and highly obscured by dust (Haas et al. 
2003).  Thus, young quasars can be effectively searched using their 
mid-infrared (MIR) spectral energy distribution (SED).  However, to avoid 
confusion with dusty starbursts, whose MIR SED can be similarly red, an 
additional diagnostic besides the MIR color is needed to identify AGN-powered 
sources.  
Here we use the presence of a strong, compact radio source to confirm the AGN 
nature of our mid-infrared selected objects, in a similar fashion to 
Mart\'{i}nez-Sansigre et al. (2005; Nature, 436, 666).
Our goal is to see whether {\it young}\ 
radio quasars show signatures of outflows that might be indicative of AGN 
feedback during the early stages of their evolution.

We use the {\it Wide-Field Infrared Survey Explorer}\ ({\it WISE}; Wright et 
al. 2010) to find luminous, obscured quasars (C. J. Lonsdale et al., in 
preparation).  The sample is selected to be bright in 22~$\mu$m ($S_{22~\mu {\rm m}} > 4$ mJy) and to have extremely red colors in the MIR, $1.25 
[m(3.4~\mu{\rm m})-m(4.6~\mu{\rm m})] + [m(4.6~\mu{\rm m})-m(12~\mu{\rm m})] > 
7$ in the Vega magnitude system.  Then we match the red objects from 
{\it WISE}\ to the NRAO VLA Sky Survey (Condon et al. 1998) 
to select sources 
with $S_{1.4~{\rm GHz}}/S_{22~\mu {\rm m}} > 1$ to ensure that the sample 
is powered by AGNs (Ibar et al. 2008). Finally, we only include sources with 
no extended radio structure, in order to avoid evolved radio galaxies.  This 
results in 156 sources. 
We are in the process of performing follow-up 
observations to explore the origin of these sources and to search for 
observational signatures of AGN feedback.  Here we present near-infrared (NIR)
spectroscopic data of a subset of these young luminous quasars.  Throughout 
the paper we adopt $H_0 = 71$ \kms\ Mpc$^{-1}$, $\Omega_m=0.27$, and 
$\Omega_\Lambda=0.73$.  

\section{Spectroscopic Observations}

As our sources are selected from {\it WISE}\ and are chosen to be very dusty, 
they are expected to be very faint at shorter wavelengths.  No reliable 
photometry is available in the optical or NIR.  From the observed W1 
(3.6 $\mu$m) flux density and an SED template for type 2 AGNs (Polletta et al. 
2007), we estimate that our sources are fainter than 19$-$21 AB mag in the $K$ 
band.  We obtained NIR (simultaneous $J$, $H$, and $K$) spectra for 
24 objects using the Folded-port InfraRed Echellette (FIRE) on the Baade 6.5 
meter telescope at Las Campanas Observatory on 27--29 July 2012 UT.  The data 
were taken in low-resolution prism mode using a 0\farcs6 slit. The sky 
conditions were clear and the seeing was 0\farcs8 $-$ 1\farcs6. The 
observations were taken at low airmasses 
\psfig{file=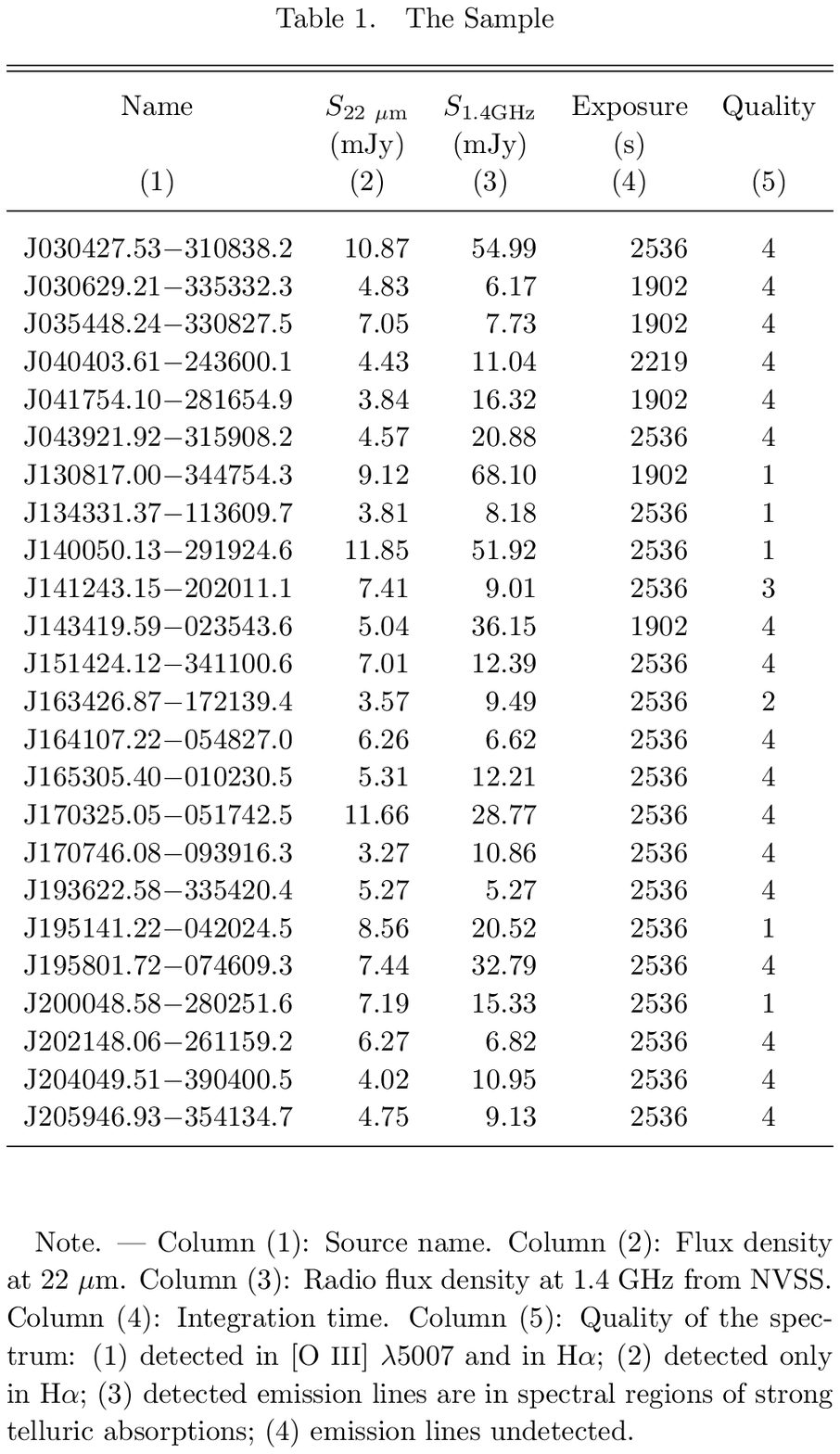,width=8.75cm,angle=0}
\vskip 0.3cm
\noindent
($\le 1.4$), with the slit aligned 
along the parallactic angle.  
Each object was dithered along the slit, with a 
position offset of 9\arcsec\ and an exposure time of 158~s in each position.  
The total integration times range from 1902~s to 2536~s, split into several 
repeats of ABBA sequences.  Following each science target, we observed a 
nearby A0V star (from a catalog provided by the Observatories) for telluric 
and relative flux calibration.  Since a majority of the selected A0V stars are 
fairly bright, we slightly offset the slit from the center of the stars to 
avoid saturation.  Absolute flux calibration was performed using two fainter, 
unsaturated stars that were properly centered on the slit.  Since all of the 
science objects are too faint to be seen on the slit image, we had to apply a 
blind offset from the nearest Two-Micron All-Sky Survey (2MASS; Skrutskie et 
al.  2006) source in order to place the object on the slit. For the typical 
brightnesses of our sample, the relative astrometric 
uncertainty\footnote{http://wise2.ipac.caltech.edu/docs/release/allsky/}
between {\it WISE}\ and 2MASS is $\sim$0\farcs4.  
Given the seeing conditions 
during the observing run, our absolute flux scale may have been underestimated 
by up to $70\%$.  Moreover, it is possible that our source blind offset 
acquisition procedure may have entirely missed some of the objects; this may 
partly account for the low detection rate reported below. 
The FIRE spectra cover 0.8 to 2.6 $\mu$m with an instrumental resolution
of $\lambda/\Delta \lambda \approx 250$ (1200 \kms), as estimated from the 
full width at half maximum (FWHM) of the arc lines. 

We used the IDL package {\tt FIREHOSE} for data reduction. Flat-field
 correction was done using images taken with an external quartz lamp. We used 
He+Ne+Ar arc lamp spectra for wavelength calibration. After extracting the 
one-dimensional spectra, we corrected for telluric absorption and applied 
flux calibration using the {\tt xtellcor} package within the Spextool 
pipeline (Cushing et al. 2004; see Vacca et al. 2003 for a detailed 
description). 
\begin{figure*}[t]
\psfig{file=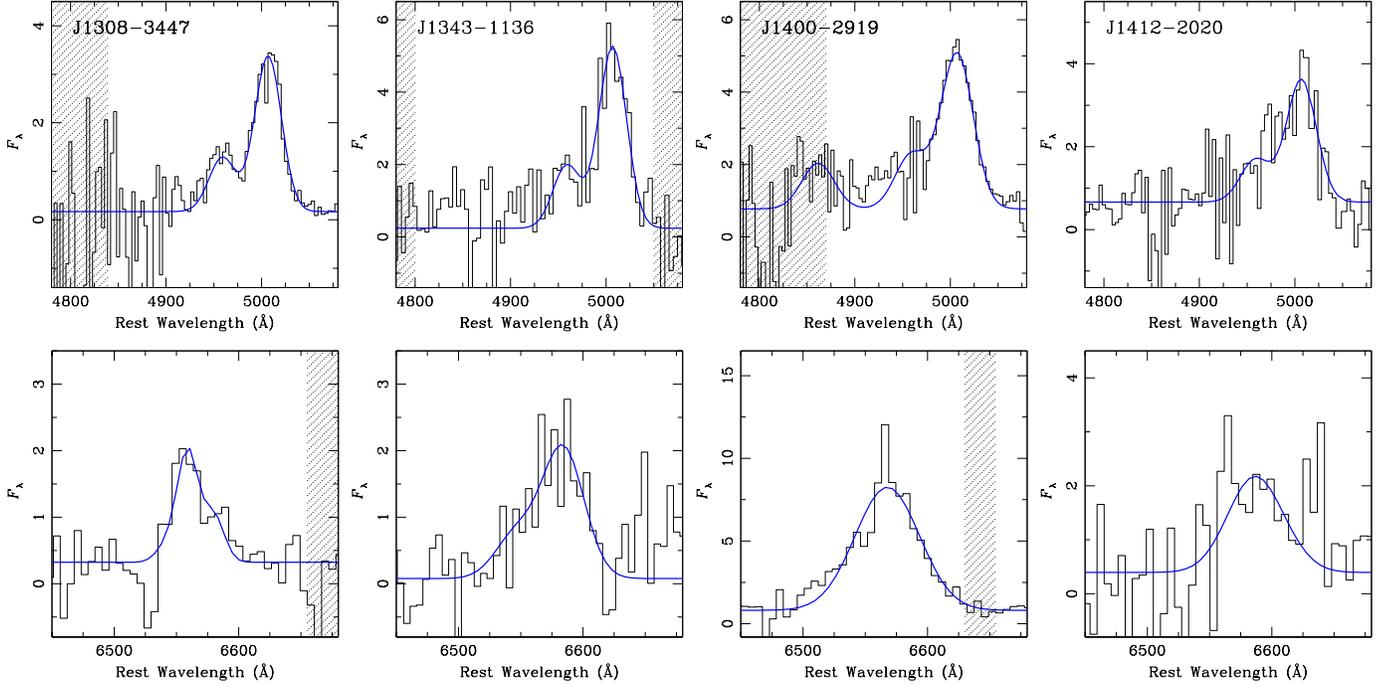,width=19cm,angle=270}
\figcaption{
FIRE spectra in the rest frame over the region 
surrounding \hb+\oiii\ and \ha+\nii. The flux density is in units of 
$10^{-16}$ \flamb. 
The original spectrum and the model for the emission lines are denoted 
by black histogram and a thick blue line, respectively. The position
of bright night sky lines is denoted by the shaded region. 
}
\end{figure*}
\begin{figure*}
\psfig{file=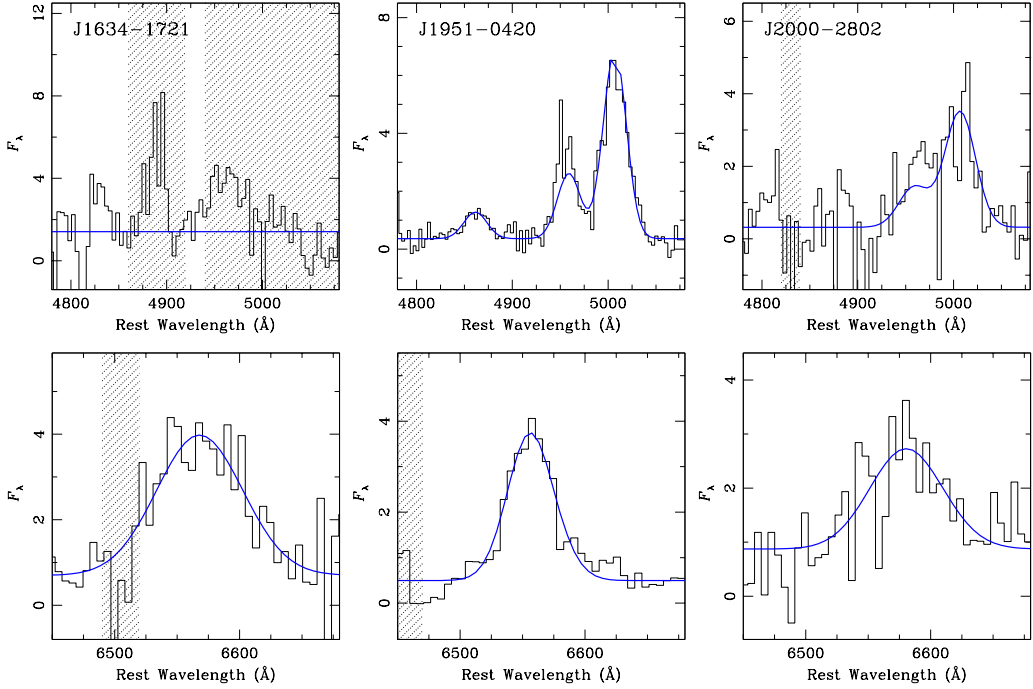,width=14.2cm,angle=270}
\figcaption{
Same as Fig. 1.
}
\end{figure*}

\begin{figure*}[t]
\psfig{file=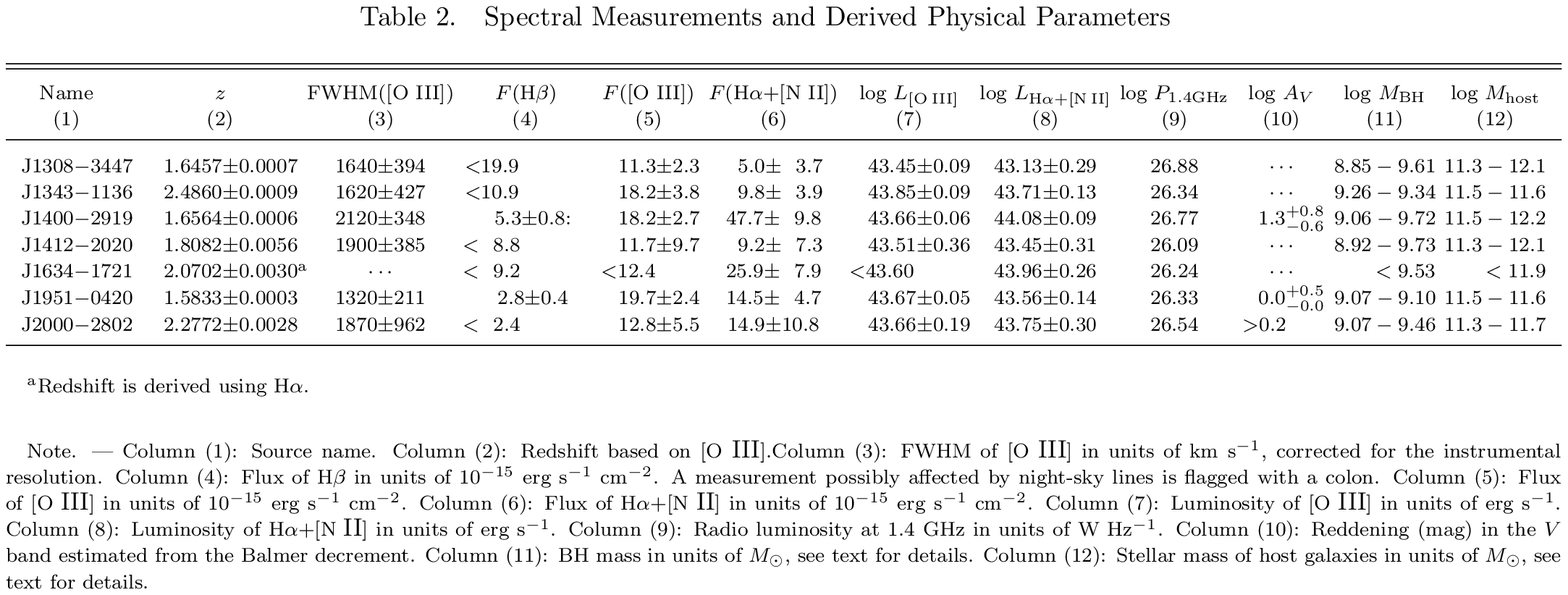,width=19cm,angle=270}
\end{figure*}

\section{Spectral Properties}
Table 1 lists the basic properties of the sample, including the total exposure 
times and the quality of the observed spectra.  The signal-to-noise ratio of 
the FIRE spectra is too low to see the underlying continuum.  We positively 
detected identifiable emission lines in seven out of the 24 objects observed 
(Table 2); their spectra are shown in Figure~1 and 2.  
The redshifts lie in the range 
$z \approx 1.6 - 2.5$.  Six show 
\oiii\ \ll4959, 5007 and \ha, but in J$1412-2020$ the lines lie in 
spectral regions of strong telluric absorption and thus are somewhat 
uncertain.  In J$1634-1721$ only \ha\ is detected.  It is difficult to know 
why the majority of the sample was undetected.  Possible reasons include: (1) 
the narrow-line region is heavily reddened by dust; (2) the emission lines 
coincide with strong OH lines or sky absorption; or (3) the slit was misplaced 
due to the astrometric uncertainty between {\it WISE}\ and 2MASS.  
In particular, since the accuracy of the telescope offsets is not precisely 
known throughout the night, it is hard to know which factors are dominantly 
responsible for the non-detection of emission lines. The detected 
sources are essentially indistinguishable from the undetected ones in terms of 
MIR flux, MIR color, or radio brightness.

We do not detect broad \hb\ or \ha\ in any of the objects.  Assuming FWHM = 
5000 \kms\ for a hypothetical broad component and \mbh\ $=10^9$ \solmass, we 
can use the \ha\ virial mass formalism of Greene \& Ho (2005b) to estimate the 
flux of broad \ha\ that should be present if our objects are genuinely 
unobscured  type 1 AGNs.  Comparing these expected values with the observed 
upper limits of broad \ha\ flux, we conclude that the broad-line region, if 
present, must be reddened by $E(B-V) \ge 1.2$ mag.  This suggests that our 
objects are true type 2 quasars rather than the lightly reddened quasars 
usually selected using NIR colors (e.g., Glikman et al.  2012).  

We follow the procedures outlined in Ho \& Kim (2009) to fit the spectra to 
obtain basic parameters for the emission lines.  We use a single Gaussian to 
fit \hb\ and each component of the doublet of \oiii\ \ll4959, 5007. The 
width is fixed to be the same for all three lines. We constrain the peak 
ratio of \oiii\ \lamb4959 to \oiii\ \lamb5007 to 2.88 and fix the separation 
between the two lines to the known value.  For objects with no detection of 
\hb, we estimate a $3\sigma$ upper limit of its flux by using the width of 
\oiii. For J1634$-$1721, where \oiii\ is also undetected, we assume a Gaussian 
profile with $\sigma = 500$ \kms.  For \ha\ and \nii\ \ll6548, 6583,
we simply measure the integrated flux of the lines because they are severely 
blended at our spectral resolution.
 
Although \hb\ is reliably detected in only two sources (J1400$-$2919 and 
J1951$-$0420), the relatively high flux ratios of \oiii\ \lamb5007 to \hb\ 
(3.4 and 7.0, respectively) suggest that the primary ionizing source of our 
sample is associated with AGNs rather than young stars. This finding is 
consistent with the strong radio emission relative to 22 $\mu$m flux.  
However, because we do not resolve \ha\ from \nii, we cannot use the \nii/\ha\ 
ratio and traditional optical line ratio diagnostic diagrams (e.g., Ho 2008) 
to completely rule out the possibility that star formation also
contributes to the ionizing continuum. For the same reason, it is also hard to 
properly constrain the extinction of the host galaxy from the Balmer decrement. 
To estimate the \ha\ flux from the total flux of \ha+\nii, we adopt an average
value of \nii/\ha\ ($0.83\pm0.42$) derived from quasars selected from the
Sloan Digital Sky Survey (SDSS; Shen et al. 2011). We estimate $A_V$ = 1.3 and 
0.0 mag for J1400$-$2919 and J1951$-$0420, respectively; the upper limit of 
\hb\ for J2000$-$2802 yields a lower limit of $A_V=0.2$ mag. These values of 
$A_V$ are surprisingly low for MIR-selected sources, but they might be biased 
in view of the large number of nondetections, which may be heavily extincted.

Interestingly and quite surprisingly, the \oiii\ lines are spectrally 
resolved for all 
detected sources. The FWHM of the line, corrected for instrumental resolution,
ranges from 1320 to 2120 \kms. 

\section{Discussion}
\subsection{Black Hole Mass and Host Galaxy Mass}

Since the bolometric luminosity of an AGN should not exceed the Eddington limit 
for a given BH mass (but see Abramowicz et al. 1988), we can estimate a
{\it lower limit}\ to the BH mass from the observed luminosity, using, as 
commonly done, the \oiii\ luminosity as a proxy for the bolometric luminosity.
The \oiii\ luminosities of 
our sample range between $L_{\rm [O~III]} \approx 10^{43}$ and $10^{44}$ \lum.
These are likely conservative lower limits because we have not accounted for 
extinction, and, in any case, our observations may suffer from significant 
slit losses (see Section 2). Yet, these \oiii\ luminosities lie in the upper 
end of the luminosity distribution of optically selected $z < 0.8$ type 2 
quasars (Reyes et al. 2008).  Using a bolometric correction factor of 
$L_{\rm bol}/L_{\rm [O~III]} = 3200$ (Shen et al. 2011), we obtain \mbh\ 
$\ge 10^{8.9} - 10^{9.3}$ \solmass. A caveat is that the correction factor
can be slightly overestimated in the radio-loud AGNs due to the 
enhanced \oiii\ (e.g., Stern \& Laor et al. 2012). 
From the correlation between BH mass and host galaxy mass, we 
suspect that these systems may be hosted by very massive galaxies at 
this epoch.  
Taking into account the cosmic evolution of the BH mass-stellar mass relation 
(Decarli et al. 2010), we estimate 
$M_{\rm host} \ge 10^{11.3}-10^{11.5}$ 
\solmass. These mass estimates are, of course, highly uncertain, 
depending on, among other things, the strict validity of the Eddington limit, 
the bolometric correction for \oiii, reddening and slit loss corrections, 
and the $M_{\rm BH}-M_{\rm host}$ relation and its uncertain redshift 
dependence.

On the other hand, we can independently estimate {\it upper limits}\ to the
mass of the hosts using the observed MIR luminosity.
Given the redshift range of the sample ($z \approx 1.6-2.5$), 
the 3.4~$\mu$m band roughly corresponds to the rest-frame $J$ band.  Since the 
contribution from hot dust emission heated by the AGN might not be negligible 
in this band (e.g., Urrutia et al. 2012), it only allows us to derive an upper 
limit to the stellar mass of the hosts.  We convert the observed 
3.4~$\mu$m flux to the rest-frame $J$ band using the template spectrum of 
type 2 Seyferts from Polletta et al. (2007).  Different choices in templates 
(e.g., starbursts instead of Seyfert 2s) lead to only moderate 
(\lax 0.1 mag) differences in the resulting $J$-band magnitudes.  
For simplicity, we adopt spectral
models from  Bruzual \& Charlot (2003) by assuming a single stellar population 
with formation redshift $z_f=$ 2.5, 3, 3.5, and 5, solar metallicity,
and a Salpeter (1955) stellar initial mass function.   With these assumptions,
we derive $M/L_J$ as a function of redshift; it varies between 0.4 and 1.  The 
average scatter in $M/L_J$ at a given redshift due to different $z_f$ is $\sim 0.1$ 
dex.  Thus, $M_{\rm host} < 
10^{11.6}-10^{12.2}$ \solmass\ and correspondingly $M_{\rm BH} 
< 10^{9.1}-10^{9.7}$ \solmass.
These estimates neglect the effects of dust extinction.  

The above considerations suggest that the host galaxies of our 
young quasars have stellar masses $M_{\rm host} \approx 10^{11.3} 
- 10^{12.2}$ \solmass.  They populate the massive end of the galaxy stellar 
mass function at $z \approx 2$ (Daddi et al. 2004) and will evolve to become 
the most massive early-type galaxies at $z=0$.  

The width of \oiii\ can serve as a surrogate for the stellar velocity 
dispersion ($\sigma_*$; e.g., Nelson 2000; Greene \& Ho 2005a; Ho 2009).  
The derived $\sigma_*$ range from 550 to 890 \kms. 
These are clearly implausible: no galaxy is known to exceed 
$\sigma_* \approx 450$ \kms\ (Salviander et al. 2008).  The \oiii\ line must 
be significantly broadened by non-gravitational forces, presumably associated 
with an outflow or disk-wind (e.g., Whittle 1992; Greene \& Ho 2005a).

\begin{figure*}[t]
\psfig{file=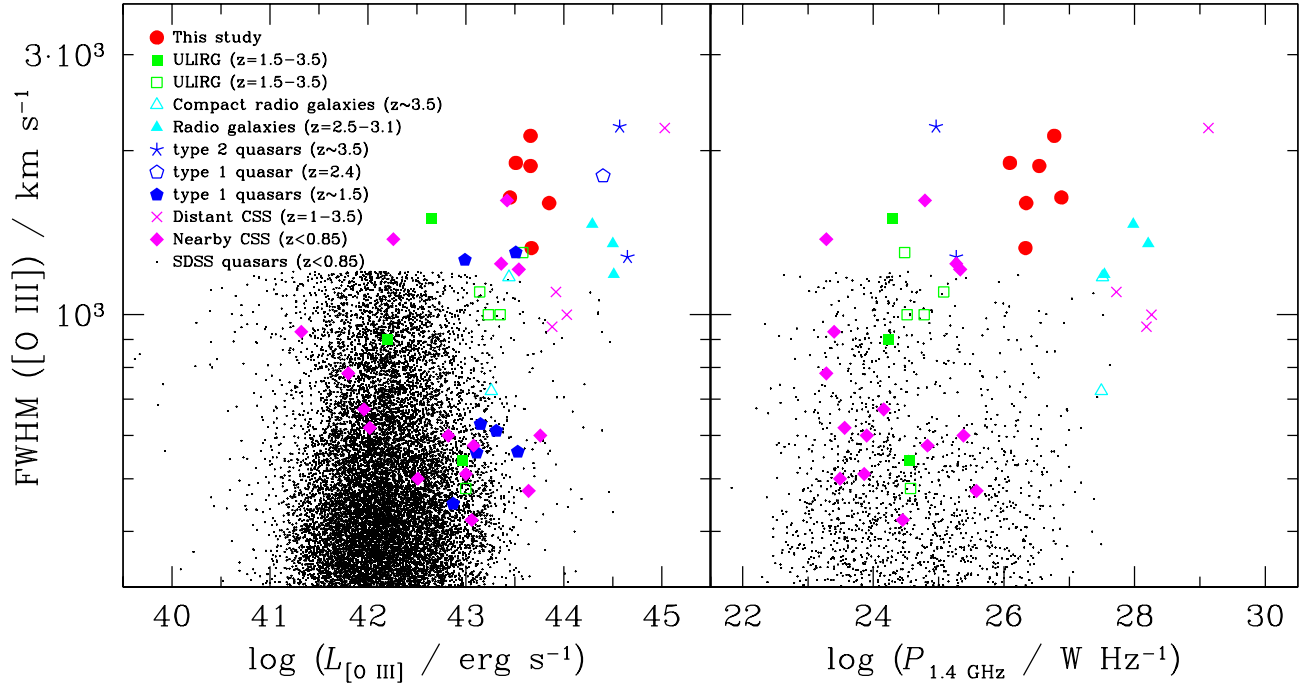,width=19cm,angle=0}
\figcaption{
({\it left})  FWHM of \oiii\ vs. \oiii\ luminosity for our sample 
({\it red circles}) and various types of AGNs: 
high-$z$ ULIRGs ({\it green squares}; Harrison et al. 2012),
high-$z$ radio galaxies ({\it cyan triangles}; Nesvadba et al. 2007, 2008),
high-$z$ type 2 quasars ({\it blue solid stars}; Nesvadba et al. 2011),
high-$z$ type 1 quasars ({\it blue pentagons}; Ho et al. 2012; 
Cano-D\'{i}az et al. 2012),
high-$z$ compact steep-spectrum (CSS) sources ({\it magenta crosses}; 
Hirst et al. 2003),
nearby CSS sources ({\it magenta diamonds}; Gelderman \& Whittle 1994),
and nearby type 1 quasars from SDSS ({\it small black dots}; Shen et al. 2011).
The FWHMs are derived either from single-Gaussian 
({\it filled symbols}) or double-Gaussian ({\it open symbols}) profiles.
For the open symbols, the FWHMs of the broad component are shown, 
while the \oiii\ 
luminosity includes both the narrow and broad components.
({\it right}) FWHM of \oiii\ vs. radio luminosity at 1.4~GHz. We assume
$f_{\nu}\propto\nu^{-0.5}$ in order to estimate the radio luminosity in the rest-frame 1.4~GHz; the overall trend does not change if the sources have a 
flat spectrum ($f_{\nu}\propto\nu^0$).
}
\end{figure*}

\subsection{AGN Feedback}
There is mounting evidence that broad \oiii\ emission is related to AGN-driven
outflows (e.g., Nesvadba et al. 2007; Harrison et al. 2012). Figure 3 (left) 
shows the correlation between the FWHM and luminosity of \oiii\ for various 
types of quasars and radio galaxies.  Where the line was fit with two 
Gaussians in the literature, we plot the broad and narrow components open and 
filled symbols, respectively; single-Gaussian fits are plotted with 
solid symbols.  Clearly, our sample has extremely broad \oiii\ line widths 
compared to other sources with similar redshift and luminosity.  
Excluding the SDSS quasars\footnote{We include the low-$z$ 
SDSS quasars for illustration purposes only. The \oiii\ FWHM for the 
sample (Shen et al. 2011), may be somewhat biased toward lower 
values because they were derived from a narrow component, while
the fit was done with two (narrow+broad) components.},
we find a weak 
correlation between FWHM and $L_{\rm [O~III]}$ 
with a Spearman's rank correlation coefficient of 0.54 
($P=0.0002$). This correlation has been known for various types of AGNs (e.g., 
Ho et al. 2003, and references therein), although our trend is driven by 
sources with extreme line widths (\gax 1000 \kms) compared to those in 
previous studies (a few hundred \kms). While the origin of the correlation 
between FWHM and $L_{\rm [O~III]}$ is still unclear (e.g., 
Whittle 1992; Ho et al. 2003), the very large line widths seen in our 
sample almost certainly do not track gravitational motions (see \S{4.1}). 
Instead, the most natural explanation for the supervirial velocities is that 
they reflect the kinematics of AGN-driven outflows, which seem to be prevalent 
in our sample of young radio quasars. 
The flux ratio of radio to \oiii\ for our sources is 
slightly lower than in high-$z$ radio galaxies and compact steep-spectrum 
(CSS) sources, indicating that our sample is moderately radio-loud.

A jet-induced origin for the large velocities is supported by the distribution 
of \oiii\ FWHM and radio power (Figure 3, right), which qualitatively 
resembles trends noted by Heckman et al. (1981) and Nesvadba et al. (2011). 
Integral-field spectroscopy of high-z AGNs with broad \oiii\ lines
finds that the emission is often associated with extended outflows of ionized
gas (e.g., Nesvadba et al. 2011; Cano-D\'{i}az et al. 2012).  The energy 
deposition of the outflows cannot be fully explained by star formation, 
suggesting that an additional source of energy must come from the AGN
(e.g., Harrison et al. 2012). Spatially resolved spectroscopy is needed to 
estimate the energy contribution from AGNs in our sample.

Our {\it WISE}-selected, radio-emitting AGNs bears an interesting 
similarity to CSS sources, whose small physical 
extent ($\sim$ a few kpc) and steep radio spectrum suggest that they are young 
radio galaxies (Fanti et al. 1990).  The broad and complex velocity profiles of 
\oiii\ observed in CSS sources are thought to arise from interaction between 
the radio jet and the interstellar medium of the host (Gelderman \& 
Whittle 1994; Holt et al. 2008).  Figure 3 shows that bona fide CSS sources 
tend to have larger \oiii\ widths compared to ordinary SDSS quasars; the line 
width enhancement is largest for more distant systems.  Our sample, 
selected to be both dusty and compact in radio morphology, qualitatively 
resembles CSS sources, although we presently do not yet have sufficient 
radio spectral information to firmly confirm this association.  Follow-up 
interferometric radio observations are under way.

Theoretical models of galaxy formation that invoke AGN feedback often link 
ULIRGs and quasars through an evolutionary sequence mediated by galaxy merging 
(e.g., Di Matteo et al. 2005).  In this model, a massive, gas-rich merger 
begins with a ULIRG phase dominated by star formation, which then evolves into
ULIRGs that host AGNs, young quasars, and then finally to optically revealed 
quasars with suppressed star formation. The strength of AGN feedback is 
expected to dramatically increase in the later stages.  In ULIRGs with AGNs, 
Harrison et al. (2012) found striking evidence for AGN feedback from the 
detection of spatially extended \oiii\ emission and broad line widths ranging 
from 900 to 1400 \kms.  Their sample is also detected in the submillimeter, 
indicating substantial ongoing star formation. Interestingly, our sample has 
even larger line widths than theirs, although the stellar masses of the host 
galaxies we deduced for our sources are comparable (e.g., Hickox et al. 2012).

\acknowledgements
We thank Robert Simcoe and Rik Williams for helpful advice on the Magellan 
observations.  An anonymous referee offered valuable input. This work is 
supported by a KASI-Carnegie Fellowship (MK) and 
the Carnegie Institution for Science (LCH).

\end{document}